# Sub-Nyquist Field Trial Using Time-Frequency-Packed DP-QPSK Super-Channel Within Fixed ITU-T Grid

L. Potì, G. Meloni, G. Berrettini, F. Fresi, T. Foggi, M. Secondini, L. Giorgi, F. Cavaliere, S. Hackett, A. Petronio, P. Nibbs, R. Forgan, A. Leong, R. Masciulli, C. Pfander

*Abstract*— Sub-Nyquist time frequency packing technique was demonstrated for the first time in a super-channel field trial transmission over long-haul distances. The technique allows a limited spectral occupancy even with low order modulation formats. The transmission was successfully performed on a deployed Australian link between Sydney and Melbourne which included 995 km of uncompensated SMF with coexistent traffic. 40 and 100 Gb/s co-propagating channels were transmitted together with the super-channel in a 50 GHz ITU-T grid without additional penalty. The super-channel consisted of eight sub-channels with low-level modulation format, i.e. DP-QPSK, guaranteeing better OSNR robustness and reduced complexity with respect to higher order formats. At the receiver side, coherent detection was used together with iterative maximum-a-posteriori (MAP) detection and decoding. A 975 Gb/s DP-QPSK super-channel was successfully transmitted between Sydney and Melbourne within four 50GHz WSS channels (200 GHz). A maximum potential SE of 5.58 bit/s/Hz was achieved with an OSNR=15.8 dB, comparable to the OSNR of the installed 100 Gb/s channels. The system reliability was proven through long term measurements. In addition, by closing the link in a loop back configuration, a potential SE·d product of 9254 bit/s/Hz·km was achieved.

## I. INTRODUCTION

IN order to fulfill market demand, optical transport networks need to improve in term of flexibility, power consumption, and capacity. The capability to support increasing spectral efficiency (SE) on installed fiber is critical in allowing carriers to leverage their existing investments. Several techniques have been proposed to increase the SE such as Nyquist wavelength division multiplexing (WDM) [1], sub-Nyquist channel spacing transmission [2], orthogonal frequency division multiplexing (OFDM) [3], and time-frequency-packing (TFP) [4]-[5]. Both Nyquist-WDM and OFDM solutions, whose performance and complexity are basically equivalent on the fiber-optic channel [6], employ orthogonal signaling. In these two cases, the orthogonality condition sets a lower limit to time- and frequency-spacing (the Nyquist criterion), such that the achievable spectral efficiency is limited by the number of levels of the underlying modulation format. Higher spectral efficiency requires higher-level modulation, e.g., 16-quadrature amplitude modulation (QAM), with higher complexity and lower resilience to signal-to-noise ratio (SNR) degradation and nonlinear effects. As a theoretical upper limit in terms of SE, the Nyquist-WDM approach allows a channel spacing equal to the baud-rate, avoiding both cross-talk due to inter carrier interference (ICI) and inter-symbol interference (ISI). This technique consists of transmitting channels with a rectangular spectrum and a bandwidth equal to the baud-rate. This results in channels at different wavelengths not overlapping (ICI is avoided) and ISI does not occur at the optimum sampling instant due to the Sinc-shaped pulses in the time domain. However, the concept of Nyquist-WDM would require the use of a specific optical spectral shaping, i.e., specific optical filters, as discussed in [7]. Typically, fourth-order super-Gaussian optical filters are used rather than ideal rectangular ones and small guard bands must be considered, thus reducing the theoretical achievable SE. Alternatively, Nyquist-WDM can also be obtained by electrically shaping the modulator driving signals through analogue filters or digital signal processing (DSP) by means of high speed digital-to-analog converters (DACs) at the transmitter side. In the former case, a quasi-rectangular shape is obtained by accepting a bandwidth increase between 20-25%, thus reducing the SE. In the latter case, the complexity and cost of the transmitter is increased due to the requirements in terms of electrical bandwidth and sampling rate for the DAC. The above mentioned techniques cannot provide an ideal rectangular shape due to technology limitations [8]. OFDM provides spectral efficiency performances comparable to Nyquist-WDM techniques, but at the expense of more complex DSP; as extensively detailed in [6]. In long-haul optical links the advantages offered by this technique (e.g. bit and power loading, pulse shaping) cannot be easily exploited, namely due to technological constraints and to the flat frequency response of the optical channel in the linear propagation regime.

Recently, the TFP approach has been proposed which, giving up the orthogonality condition, can overcome the Nyquist limit and achieve higher SE with low-level modulations [4], [9], [10]. This approach is an extension of faster-than-Nyquist (FTN) signaling [11]. By increasing signaling rate for a fixed bandwidth beyond the Nyquist limit, some bandwidth is saved at the expense of introducing ISI. In FTN signaling, pulses can be packed closer than the Nyquist limit without performance degradation, provided that the minimum Euclidean distance of the system is not reduced and the optimum detector is employed (Mazo limit) [11]. This approach, however, does not provide the best performance in terms of SE and poses no limits to the complexity of the required detector.

Alternatively, TFP overcomes these limitations and seeks the best solution in terms of SE for a fixed detector complexity. In particular, time packing consists of finding the optimum time spacing between pulses, which, despite introducing ISI, maximizes the achievable SE for a given input constellation and detection strategy. The same concept is extended to the optimization of frequency spacing in TFP. By maximizing the SE for a given modulation format, TFP also assures the lowest requirements in terms of electrical bandwidth both at the transmitter and receiver and, as demonstrated by this field trial, full compatibility with existing wavelengths with the capability of extending current installed infrastructure beyond 100 Gb/s with minimal need for infrastructure renewal.

In order to increase link capacity and distance, orthogonal techniques have been used in experimental demonstrations in combination with optical space, time, polarization, and wavelength division multiplexing (SDM, OTDM, PDM, WDM). In particular, the highest symbol rate of 160 Gbaud exploiting OTDM and the largest single channel bit rate of 768 Gbit/s using dual polarization (DP)-64QAM have been demonstrated in [11] and [13], respectively. In terms of system capacity, a record value of 105 Tb/s has been demonstrated in [14] using multi-core fiber. In [15] the authors experimentally demonstrated transmission of a 4 Tb/s super-channel over a 12000 km link with time-domain hybrid QPSK-8QAM modulation format.

Moreover, the highest spectral efficiency of 32 bit/s/Hz has been obtained through PDM-16QAM coherent transmission over 12 spatial and polarization modes of a few-mode fiber [16]. Among field trial demonstrations, in [17] they demonstrated 21.7 Tb/s using a DP-8QAM/QPSK modulation format over 1503 km with a maximum SE of 5.26 b/s/Hz while in [18] they demonstrated DP-8QAM and DP-16QAM transmission over real installed links of 1822 km and 634 km respectively, with high total capacity and a SE×distance ($d$) product of about 9000 bit/s/Hz·km.

In this paper, we successfully demonstrate TFP transmission over an uncompensated installed link between Sydney and Melbourne operated by Telstra including 100 GHz spaced 40 and 100 Gb/s co-propagating traffic. Resorting to information theoretic principles and advanced signal processing techniques, we optimized sub-channel bandwidths and inter-channel spacing, maximizing the SE and achieving a SE·$d$ record for field trial demonstration of 9254 bit/s/Hz·km with an optical SNR (OSNR) of 14.8 dB. In addition, a 975 Gb/s DP-QPSK super-channel was successfully transmitted in compliance with the 50 GHz International Telecommunication Union – Telecommunication Standardization Bureau (ITU-T) grid over single mode fibers (SMFs) and one installed ROADM within 200 GHz of spectral occupancy, with an OSNR=15.6 dB. A maximum potential SE of 5.58 bit/s/Hz was demonstrated with an OSNR of 15.8 dB. In order to verify the system robustness performance was measured including artificial PMD addition and long term measurements confirming the system reliability.

The paper is organized as follows. Section II describes TFP principles of operation giving an example of a possible practical transceiver implementation. In section III the details of the implemented system and the installed link are described. Section IV collects field trial detailed results including single way transmission, long-term measurements, and loopback transmission.

## II. TIME-FREQUENCY-PACKING

This section provides an introduction to TFP. For the sake of simplicity, only time packing is described, though the same concept can be extended also to frequency separation between channels.

A binary signal (e.g., one of the four quadratures of a DP-QPSK signal) is obtained through a sequence of pulses with band-pass bandwidth $B$, given shape, amplitude '1' or '-1', and symbol time, $T = 1/R$, where $R$ is the signaling rate. According to the Nyquist criterion, ISI can be avoided by using proper pulse shapes only if $R \leq B$, with the equal sign holding for ideal Sinc pulses. In this case, the optimum receiver is a symbol-by-symbol receiver. According to the FTN technique, it is possible to increase the signaling rate beyond the Nyquist limit, up to a certain rate $\acute{R} > B$ (Mazo limit), so that the minimum Euclidean distance between sequences (and, therefore, the symbol error probability for high signal-to-noise ratios) is preserved despite the presence of ISI, provided that the optimum sequence detector is employed [11]. However, this approach does not provide the maximum SE (the Mazo limit is $\acute{R} \cong 1.25B$ for ideal *Sinc* pulses) and poses no limits to the detector complexity (which can be extremely high).

Alternatively, the TFP includes a constraint on detector complexity and relies on information theory to determine the maximum achievable SE. When the Nyquist criterion is satisfied, the amount of information $I$ that is transferred from source to destination by each symbol (mutual information) depends only on the signal-to-noise ratio, the maximum being 1 bit/symbol (for binary modulation). Correspondingly, the amount of information that is transferred per unit time and bandwidth is $SE = I/(BT)$ and its maximum is 1 bit/s/Hz. If the signaling rate is increased beyond the Nyquist limit, the mutual information is reduced by the presence of ISI. When employing a suboptimum detector that accounts only for a limited amount of ISI, the achievable information rate $\hat{I}$ per symbol may be even lower. However, since the symbol time (at denominator of the SE) is shorter, the overall SE may increase. The time-packing approach consists of finding the signaling rate $\hat{R} = 1/\hat{T}$ which provides the maximum achievable SE, $\widehat{(SE)} = \hat{I}/(B\hat{T})$ for a given detection strategy. Even when employing low-complexity detectors, the achievable SE turns out to be much higher than when considering the Nyquist or the Mazo limit [11]. A simple example of a practical TFP transceiver is shown in Fig. 1. In order to generate a DP-QPSK modulated signal, each polarization includes two data streams whose gross transmission rate has been fixed at 35 Gb/s each.

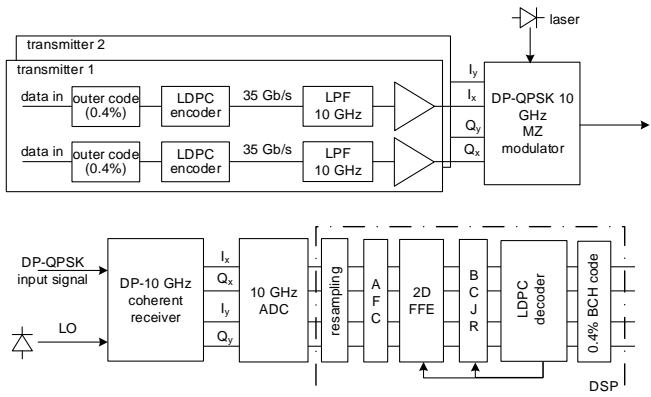

Fig. 1 TFP transmitter (top) and receiver (bottom).

The system exploits different LDPC codes to trade-off net SE and error correction capability [19]. In particular, self-designed irregular low density parity check (LDPC) codes [20] can be configured [21] with a rate of 9/10, 8/9, 5/6, 4/5, 3/4 and 2/3. A code length of 64800 bit was used. Net code gain (NCG) together with the variable node degree and the check node degree are reported in Tab. I for different code rates at a BER=$10^{-7}$ when no ISI is introduced.

By keeping the line rate constant, it is possible to adapt the optical interface therefore adjusting the transmitter capacity as a function of the propagation conditions (i.e., accumulated OSNR and propagation penalties due to fibre nonlinearity). Detection of this kind of codes is typically characterized by the presence of error floors at high SNRs. Simulations for codes in Tab. I indicate that an error floor, if present, is located at a BER lower than $10^{-8}$ [22]. In any case, outer hard-decision external codes with very low overhead can be employed to correct the residual errors and remove the floor. For instance, in the DVB-S2 standard, where LDPC codes with same length and rates as in Tab. I are adopted, outer BCH codes with less than 0.4% overhead are used to correct from 8 to 12 residual errors (depending on the rate) per codeword [23]. The transmitter assembles a total gross capacity 140 Gb/s (70 Gb/s per polarization) for a net information rate ranging between 89.6 and 120.96 Gb/s. A ninth-order Chebyshev low-pass filter with a cut-off frequency of 10 GHz (much lower than the Nyquist limit, which is 17.5 GHz) generates the TFP signal. The resulting optical bandwidth of 20 GHz gives an SE value within 4.48 and 6.05 as a function of the code rate. 10 GHz electrical low-pass filtering performed at the transmitter allows low bandwidth requirements for the electronics and the optoelectronics circuits such as the drivers and the MZ IQ modulator.

At the receiver side, a standard coherent detection scheme is used. The four analogue signals coming out from the 10 GHz photodetectors included in the receiver are digitized through an 8-bit 10 GHz ADC. Even if the symbol rate is 35 Gbaud, TFP allows the use of a reduced bandwidth ADC, whose resolution, linearity, and power consumption are better as the bandwidth gets lower. Digital signal processing includes resampling, used because of two samples/symbol, automatic frequency control (AFC) that compensates for frequency mismatch between the incoming signal and the local oscillator (LO) [24] and a two dimensional fractionally-spaced feed-forward equalizer (2D-FFE) that compensates for linear propagation impairments. The equalizer output feeds four parallel 4-state Bahl, Cocke, Jelinek, Raviv (BCJR) detectors [19], working on the four signal quadratures, followed by four LDPC decoders. The BCJR and LDPC blocks iteratively exchange information to achieve MAP detection according to the turbo-equalization principle [25]. The maximum number of iterations directly establishes the DSP intrinsic latency and was fixed to 20 even if the average number of iteration is generally lower than 10, as will be shown in the experiment description. Phase noise is treated as described in [26]. The 2D-FFE equalizer is adaptively controlled to converge to the matched filter output of the Ungerboeck observation model [27]. Equalizer update is relatively slow being based on code word basis. The estimated channel coefficients (the five central ones) are used to determine path metrics for the BCJR detectors. With respect to a traditional receiver, the added complexity is represented by the BCJR detector presence, and it is proportional to the number of states. In this case, the complexity of a 4-state BCJR detector is negligible with respect to the decoder.

TABLE I
CODE PERFORMANCES

| Code rate | NCG | Variable node degree | Check node degree |
|---|---|---|---|
| 2/3 | 11.23 | 13 | 11 |
| 3/4 | 10.03 | 12 | 15 |
| 4/5 | 9.65 | 11 | 19 |
| 5/6 | 9.15 | 13 | 22 |
| 8/9 | 8.13 | 4 | 28 |
| 9/10 | 7.91 | 4 | 31 |

III. SYSTEM DESCRIPTION

The field trial was performed over an installed fiber link, provided by Telstra, connecting the two largest Australian cities, Sydney and Melbourne, through the intermediate node in Canberra as shown in Fig. 2.

The super-channel was generated in Sydney, as detailed in *Section III B*. A single carrier coherent optical receiver was also used in Sydney both for back-to-back system characterization and loop back measures. In order to validate the system robustness with respect to the presence of traditional traffic over the same link, four DP-QPSK commercial cards (1x100 Gb/s and 3x40 Gb/s) were used in the experiments. The four cards are multiplexed together using a 100 GHz multiplexer having 50 GHz filters on the ITU-T grid. A 3 dB coupler was used in order to add the super-channel. The Melbourne site hosted 40 and 100 Gb/s coherent receivers. As shown in Fig. 2, the signal arriving in Melbourne could be either received or looped back through an almost identical link, including a wavelength selective switch (WSS) in Canberra. The signal was received in Sydney after a total propagation distance of about 1990 km.

In the following subsections, link, super-channel transmitter, and coherent receiver will be detailed.

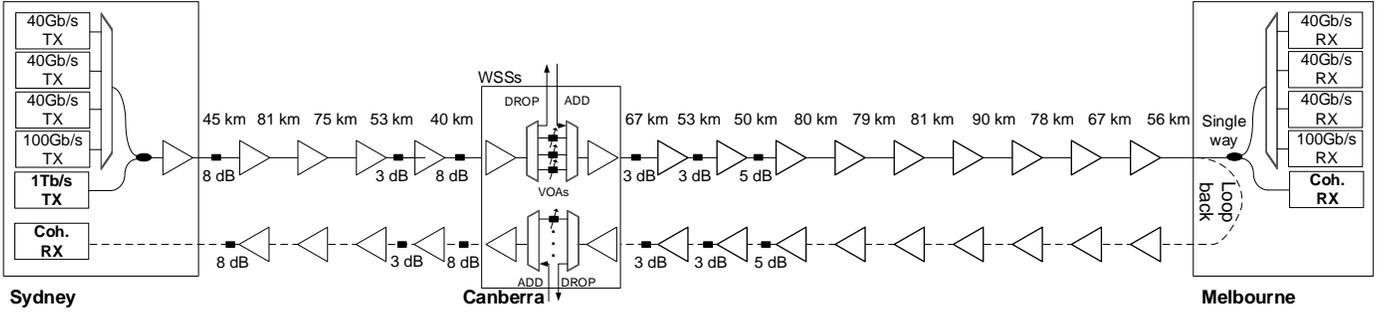

Fig. 2 Telstra fiber link between Sydney and Melbourne together with transmitter and receiver setup (the optical attenuations between amplifiers are reported as black box).

## A. Link

The 995 km link is divided into 15 uncompensated SMF spans, whose lengths are indicated in Fig. 2. Shorter spans are padded to match the minimal nominal gain value of the subsequent amplifier. Each amplifier could be configured in order to guarantee a power ranging between -3 dBm and +1 dBm for each 50 GHz channel with a limitation for the total output power of +22 dBm (i.e., 128 50GHz channels @ +1 dBm). A reconfigurable add/drop multiplexer (ROADM) site, based on a 50 GHz WSS, is present after 294 km (5 spans) in Canberra.

The WSS is composed of a 1:$N$ 50 GHz demultiplexer, $N$ variable optical attenuators (VOAs), and a $N$:1 50 GHz multiplexer, being $N$ the number of frequency slots. At each port a power monitor is used for automatic channel equalization. Even if the nominal slot bandwidth is 50 GHz for each WSS port, the real 3 dB value is generally lower. Although 40 and 100 Gb/s cards do not suffer from such a negligible optical bandwidth reduction, in the case of a super-channel an accurate characterization of the available bandwidth is mandatory in order to maximize overall SE. For this reason, amplified spontaneous emission (ASE) was used as broadband light source at the system input. All the power controls were disabled in order to prevent automatic power equalization. The measured WSS transfer function when four adjacent ports are activated is shown in Fig. 3. A power tilt of about 0.6 dB was measured between outer channels. In the considered spectral region, ASE can be assumed almost flat. In our experiments, two carriers were used for each 50 GHz slot for a total number of eight sub-carriers within 4 slots that corresponds to 200 GHz. In practice, each sub-carrier was placed at a distance of 10 GHz form the nominal central frequency of the filter.

In this way, the systems became unequally spaced where two sub-carriers occupying the same slot are 20 GHz spaced and the first sub-carrier in the adjacent slot is 30 GHz apart (markers in Fig. 3). For each slot the filter exhibited transmission asymmetries. In particular, the low frequency edge was much smoother with respect to the high frequency edge. Slightly different performances for odd and even sub-carriers were then expected, unlike a gridless scenario, where multiple carriers used for super-channel transmission can be equally spaced allowing SE maximization.

## B. Super-channel transmitter

A total aggregate gross bit rate of 1.12 Tb/s was obtained by means of eight optical carriers [4], each modulated by a 140 Gb/s narrow filtered DP-QPSK signal, corresponding to 35 Gbaud. Sub-channels' bandwidth and spacing were optimized according to the TFP technique [4][10] to maximize the achievable SE with the desired detector complexity. In addition, unequal channel spacing (20/30 GHz) was adopted to account for the guard bands between adjacent 50 GHz WDM slots, not required by TFP but imposed by the mid-link WSS (as described in the previous subsection). For that reason, a net wasted bandwidth of about 30 GHz must be considered. That translates in a SE reduction of about 15%. The experimental setup is reported in Fig. 4, showing that the odd and even channels were separately modulated by means of two integrated double nested Mach-Zehnder modulators (IQ-MZM).

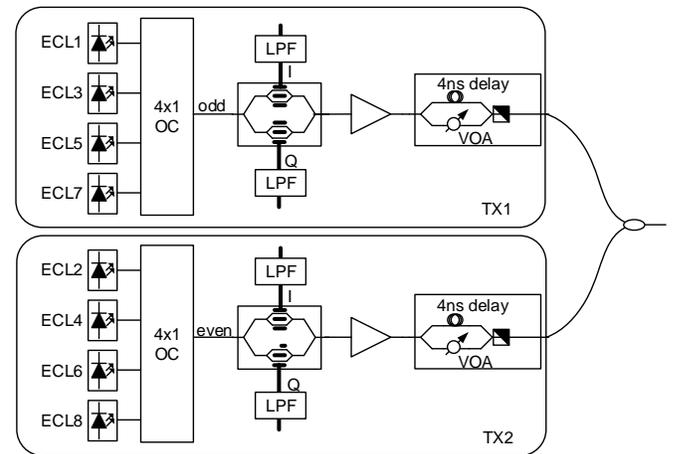

Fig. 4 TFP transmitter

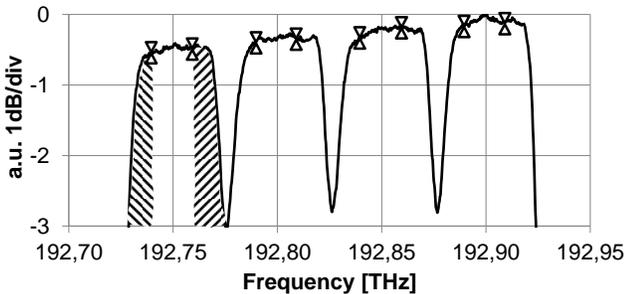

Fig. 3. Canberra WSS characterization with particular emphasis on the filter asymmetry. Markers represent carriers' wavelength.

By applying 35 Gb/s coded electrical binary signals to the in phase (I) and quadrature (Q) ports of the modulators, 70 Gb/s QPSK channels were obtained. The bit rate was then further doubled up to 140 Gb/s per channel, by emulating polarization multiplexing through a 50/50 beam splitter, a 4 ns optical delay, and a polarization beam combiner (PBC).

*C. Coherent receiver*

At the receiver side, coherent polarization-diversity detection with digital signal processing was exploited.

For each sub-channel, the receiver scheme was the one shown in the lower part of Fig. 1. However, as previously discussed, lower bandwidth and sampling rate are required by the TFP technique. The photo-detected signals at the output of four balanced photodiodes (two for each polarization) were sampled and digitized using a commercial four 8-bit, 10 GHz bandwidth analog-to-digital converter (ADC) and then digitally processed. A Tektronix DPO72004B was used to sample and store data at 25 Gs/s and bandwidth limited to 10 GHz.

Off-line digital processing was performed considering blocks of $8 \cdot 10^5$ samples at a time. Accumulated dispersion (~17000 ps/nm) was compensated using a frequency domain equalizer, while a 33 taps, two dimensional fractionally-spaced feed-forward equalizer (2D-FFE) accounted for other linear impairments (e.g., polarization rotation, residual dispersion, polarization mode dispersion (PMD)).

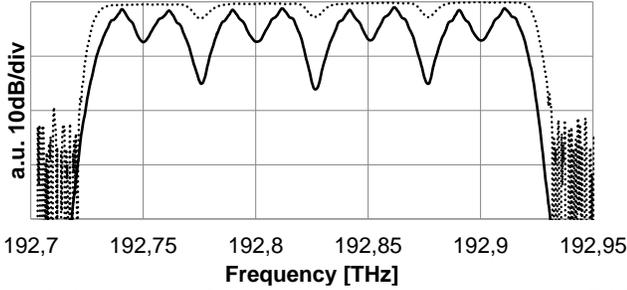

Fig. 5. Optical spectrum of the generated super-channel (solid) and the WSS profile (dotted)

## IV. FIELD TRIAL RESULTS

In order to measure the performance of the proposed techniques over the installed link, several experiments were carried out. In Fig. 5 the optical spectrum of the super-channel generated at the transmitter side in Sydney is depicted. The eight DP-QPSK channels were unevenly spaced (20/30 GHz) considering the asymmetric channels spacing required for the transmission of the super-channel through the WSS in Canberra. In particular, four pairs of channels were centered in the four 50 GHz spaced WSS ports corresponding to the central frequencies of 192.75 THz, 192.8 THz, 192.85 THz and 192.9 THz respectively. The back-to-back measurements are detailed in *Section IV A,* whereas transmission optimization and performances are described in *Section IV B*. *Section IV C* is dedicated to the exploration of SE maximization for the system having the strong constraint of 50 GHz WSS over the link. For all measurements, SE optimization is obtained through code rate adaptation.

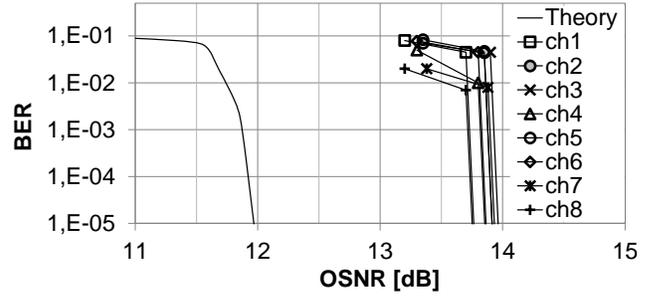

Fig. 6. Back to back measurements for all the channels. The theoretical curve is also included as a reference. ~2dBs of implementation penalty was observed.

*A. Back to back measurements*

In Fig. 6, back to back measurements performed for each channel are reported, considering a baud rate of 35 Gbaud and code rate 5/6. Reference BER measures were performed by averaging over 10 blocks of samples, each containing 8 randomly-selected code words per quadrature, per polarization, for a total of $(10 \times 8 \times 64800 \times 2 \times 2 \times R_c)$ information bits (when a code rate $R_c$ is used). BER values below $10^{-6}$ could not be reliably measured, such that the LDPC error floor (expected below $10^{-6}$) was not observed in the experimental setup. The benchmark theoretical curve is also reported. A measured implementation penalty of approximately 2 dB was measured at a BER of $10^{-5}$ for all the channels.

*B. System optimization and transmission performance*

In order to investigate the contribution of intra-super-channel non-linear effects due to propagation, the super-channel was transmitted through the link between Sydney and Melbourne (995 km) and the achieved SE was measured as a function of the channel power (Fig. 7). The measurement was performed on channel 4 (one of the central channel), over an occupied bandwidth of 200/8 GHz. Fig. 7 indicates an optimal channel power of about -1 dBm that maximizes SE, providing the best compromise in terms of ASE noise and nonlinear fiber propagation including SPM and XPM. By keeping the output EDFA power constant per sub-channel, SE optimization provided different values for each channels, depending on the OSNR and WSS filtering effect. The power penalties were measured over all the sub-channels after 995 km transmission with respect to the back to back as shown in Fig. 8. A maximum penalty of 1.3 dB is observed.

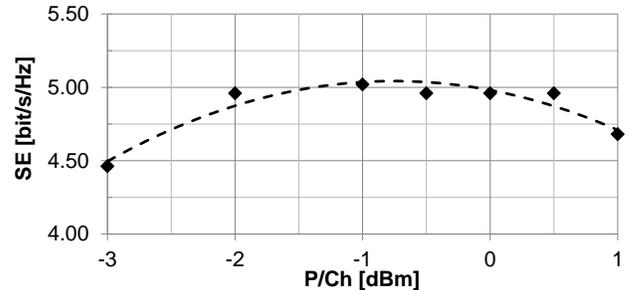

Fig. 7. Spectral efficiency as function of the power per sub-carrier

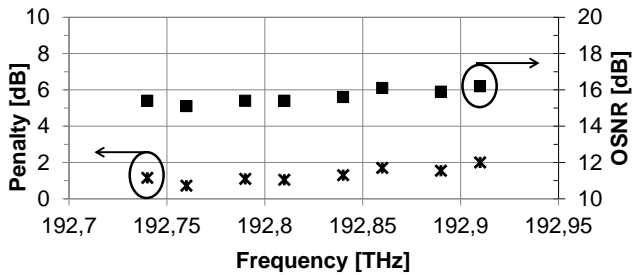

Fig. 8. Channels power penalties (left) and OSNR for a channel power of -1 dBm (right)

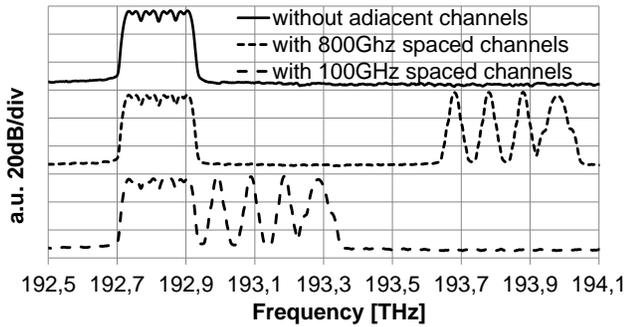

Fig. 9. Transmitted optical spectra: time-frequency-packed DP-QPSK super-channel without adjacent channels, with 40 and 100 Gb/s channels 800 GHz spaced, with 40 and 100 Gb/s channels 100 GHz spaced

The OSNR is also reported, for completeness, in the same picture, for all sub-channels, considering a channel power of -1 dBm. Taking the optimized sub-channel power as -1 dBm the received OSNR ranges between 15 and 16 dB along the channels with 0.1 nm resolution bandwidth, a value which is compatible with the majority of deployed DWDM links.

Another important aspect is the coexistence with standard traffic over the link. For that purpose, one 100 Gb/s and three 40 Gb/s DP-QPSK channels were transmitted together with the eight carrier super-channel. The initial frequency gap between the super-channel and the co-propagating channels was 800 GHz. In a second measurement, it was reduced down to 100 GHz in order to investigate fiber non-linear cross-talk effects. Fig. 9 shows the input and optical spectra for the considered configurations: a) transmitted super-channel without 40 and 100 Gb/s channels, b) transmitted super-channel with 40 and 100 Gb/s channels 800 GHz spaced, c) transmitted super-channel with 40 and 100 Gb/s channels 100 GHz spaced. The achieved SE for each channel is reported in Fig. 10.

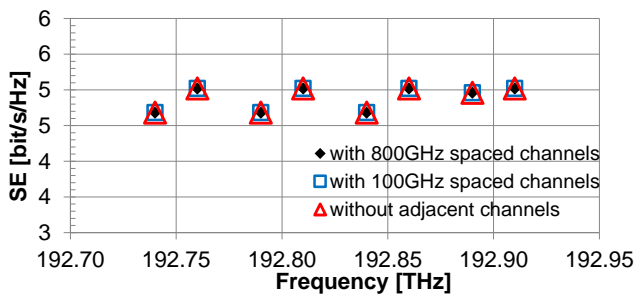

Fig. 10. Individual sub-channel maximum SE in three different cases: without adjacent channels (triangles), with 100 GHz spaced channels (squares), and with 800 GHz spaced channels (diamonds).

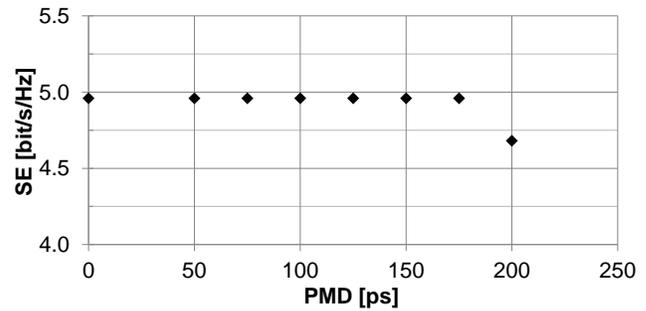

Fig. 11. SE as a function of the additional PMD

An average SE=4.87 has been obtained giving an overall information rate of 975 Gb/s. Odd channels perform slightly worse due to the asymmetric filtering response of the WSS highlighted in Fig. 3. Moreover, channel 7 slightly overperforms the other odd channels due to larger available OSNR. The three cases of Fig. 10 show no appreciable performance difference, giving evidence of negligible interference between the super channel and neighboring 40 and 100 Gb/s channels. Similarly, no penalty was measured on the 40 and 100 Gb/s channel due to the super-channel presence. In order to verify the receiver capability to compensate linear impairments, a PMD emulator was added just before the coherent receiver, introducing an additional differential group delay up to 200 ps. SE values remained unchanged introducing 170 ps, and dropped of 5% with 200 ps. Results have been measured over channel 4 and are reported in Fig. 11.

In order to validate the system stability, long term error measurements were performed. In particular, for a more accurate statistic about the system operation, new code words were generated every 15 minutes, and the number of errors computed. In Fig. 12 the average number of iterations (diamonds) are reported for all the measurements taken over 12 hours together with the number of counted errors (triangles), confirming a stable operation over the entire period. Even if the number of iteration can change statistically due to the presence of noise and to the code word, a maximum number of iterations equal to 20 was used in order to keep the latency constant.
The estimated frequency offset between the considered received channel and the local oscillator, evaluated by the AFC algorithm, is also reported in Fig. 13 for each measurement of Fig. 12.

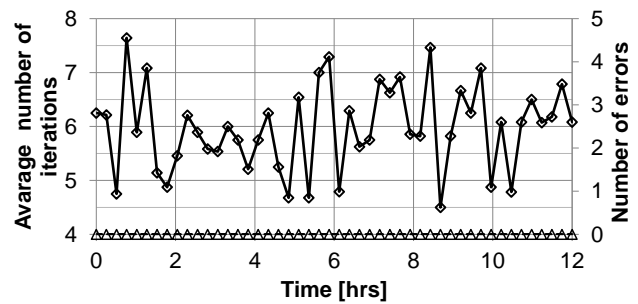

Fig. 12. Required iterations for BER computation and BER measurements

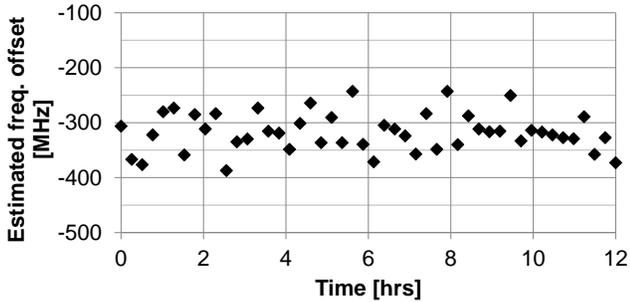

Fig. 13. Frequency offset estimated by the AFC algorithm at the receiver

*C. Three equally spaced channels experiment*

In order to verify the TFP technique potential in terms of spectral efficiency enhancement for the transmission of a DP-QPSK signal, three evenly spaced 140 Gb/s channels were transmitted through the Telstra network. The channels were 20 GHz spaced and the central channel frequency was matching with the network WSSs. The signals were first transmitted from Sydney to Melbourne (995 km-single way) and then looped back Sydney-Melbourne-Sydney (1990 km).

In both cases, the channels were filtered by the WSS in Canberra. As schematically depicted in Fig. 14, the signals were equally spaced, and the central channel suffered crosstalk from the adjacent channels, due to the shape characteristics of the electrical filter employed at the transmitter (Fig. 4).

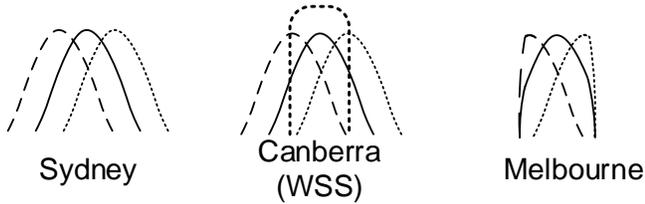

Fig. 14. Schematic of filtering effect for evenly spaced channels

The WSS at Canberra filtered out a large part of the adjacent channels, thus reducing the impact of nonlinear effects. However, the linear crosstalk among channels remained practically unchanged. The same happened for the loop back configuration (Fig. 14). For this reason, the measurement should be intended as an indicator of the potentially achievable SE (if WSS were removed), though nonlinear effects are slightly under estimated. In Table II, the potential spectral efficiency is reported together with the employed code rate and the OSNR of the received channel for the single transmission and the loop back transmission. In the loop back configuration, an SE of 4.65 bit/s/Hz was achieved, corresponding to a potential SE·d product of 9254 bit/s/Hz·km.

TABLE II
POTENTIAL SE FOR SINGLE AND LOOP BACK TRANSMISSION

| Considered scenario | Potential SE [bit/s/Hz] | Baud rate [GBd] | Employed code rate | OSNR @ RX [dB] |
|---|---|---|---|---|
| Single way | 5.58 | 35 | 4/5 | 15.8 |
| Loop back | 4.65 | 35 | 2/3 | 14.8 |

## V. CONCLUSIONS

Spectral efficiency enhancement of a Telstra installed long haul uncompensated link was proven by exploiting time frequency packing technique. A 975 Gb/s DP-QPSK super-channel was successfully transmitted through an Australian link between Sydney and Melbourne (995 km), within 200 GHz in a fixed ITU-T grid scenario. The compatibility with deployed 50 GHz grid optical filters was ensured, without network upgrading requirements. Time packing was used to maximize the achievable SE with constrained modulation and detection complexity; while LDPC rate was adapted to the available OSNR and propagation conditions to approach the achievable SE of 5 bit/s/Hz (including a reduction of 15% due to wasted guard bands in a fixed grid scenario). System performance was evaluated considering real co-propagating traffic (40 and 100 Gb/s channels) without additional penalty. The system performance was evaluated considering artificial additional PMD, verifying the receiver capability to compensate linear impairments. Long term measurements were also performed. Moreover, through loop back configuration, 1990 km transmission was demonstrated with a potential SE·d product of 9253 bit/s/Hz·km with an OSNR of 14.8 dB. The obtained results confirmed the possibility for operators to seamlessly migrate toward >100 Gb/s networks with minimal need of infrastructure renewal.